\documentclass{emulateapj}
\usepackage{graphicx}
\usepackage{amsmath}

\begin{document}
\def\etal{et al.\ \rm}
\def\ba{\begin{eqnarray}}
\def\ea{\end{eqnarray}}
\def\etal{et al.\ \rm}

\title{Ab-initio pulsar magnetosphere: three-dimensional particle-in-cell simulations of
axisymmetric pulsars}
\author{Alexander A. Philippov\altaffilmark{1} \& Anatoly Spitkovsky}

\affil{Department of Astrophysical Sciences, 
Princeton University, Peyton Hall, Ivy Lane, Princeton, NJ 08544}

\altaffiltext{1}{sashaph@princeton.edu}


\begin{abstract}

We perform ``first-principles'' relativistic particle-in-cell simulations of
aligned pulsar magnetosphere.  We allow free escape of particles from
the surface of a neutron star and continuously populate the magnetosphere with
neutral pair plasma to imitate pair production. As pair plasma supply increases,  
we observe the transition from a charge-separated
``electrosphere" solution with trapped plasma and no spin-down to a solution close
 to the ideal force-free magnetosphere with 
electromagnetically-dominated pulsar wind. We calculate the magnetospheric
structure, current distribution and spin-down power of the neutron star. 
We also discuss particle acceleration in the equatorial current sheet.

\end{abstract}


\section{Introduction}

More than forty years after their discovery, many fundamental questions about pulsars are still with us. Both the radio and gamma-ray emission lack a comprehensive theory, mainly because plasma conditions in various parts of the magnetosphere are not well understood. Production and acceleration of plasma has yet to be modeled in a self-consistent way as part of the global magnetospheric solution. 
In recent years, considerable progress has been made in constructing global models of pulsar magnetospheres under the assumption of abundant plasma supply. For example, self-consistent force-free numerical solutions of axisymmetric \citep{ckf99, gruzinov_pulsar_2005,
  mck06pulff, tim06} and oblique \citep{spit06,kc09,petri12a,lst11,
  kalap12} pulsar magnetospheres were developed. 
 These studies were extended to the full magnetohydrodynamic (MHD)
regime \citep{kom06,SashaMHD} that take plasma pressure and inertia
into account.  All such solutions generically find thin return current layers at the boundary of closed and open field lines and a strong current sheet beyond the light cylinder. Strong current regions have been identified as potential sites of particle acceleration and gamma-ray emission, mainly based on geometrical considerations of light-curve fitting \citep[e.g.,][]{Anatolygamma, greekgamma}, supported by plausible estimates of characteristic emission from current sheet reconnection \citep{ArkaDubus13, UzdenskySpitkovsky13}. 
However, 
plasma instabilities and particle acceleration in the magnetosphere that are required for ultimate emission modeling cannot be fully addressed within MHD or force-free approach, necessitating a kinetic
treatment.  The current sheet also cannot be easily studied in isolation from the rest of the magnetosphere 
as it is causally connected to the global solution. 

The goal of this Letter, therefore, is to lay the foundation for global kinetic simulations of pulsar magnetospheres. We do this by formulating a novel boundary condition for conducting rotating stars,  and use it to construct a first-principles relativistic particle-in-cell (PIC)
model of the aligned pulsar magnetosphere. We consider two extreme
limits of plasma supply: charge-separated emission from the stellar
surface in the absence of pair production, and abundant neutral plasma
injection approximating efficient pair formation. In section
\ref{sec:numerical-models}, we describe our numerical method and test
problem setup, and in section \ref{sec:pulsar} we present our results
on three-dimensional (3D) magnetospheric structure and discuss particle acceleration in the current sheet.

\section{Numerical method and setup}
\begin{figure*}

\centering
{
\includegraphics[width=0.43\textwidth]{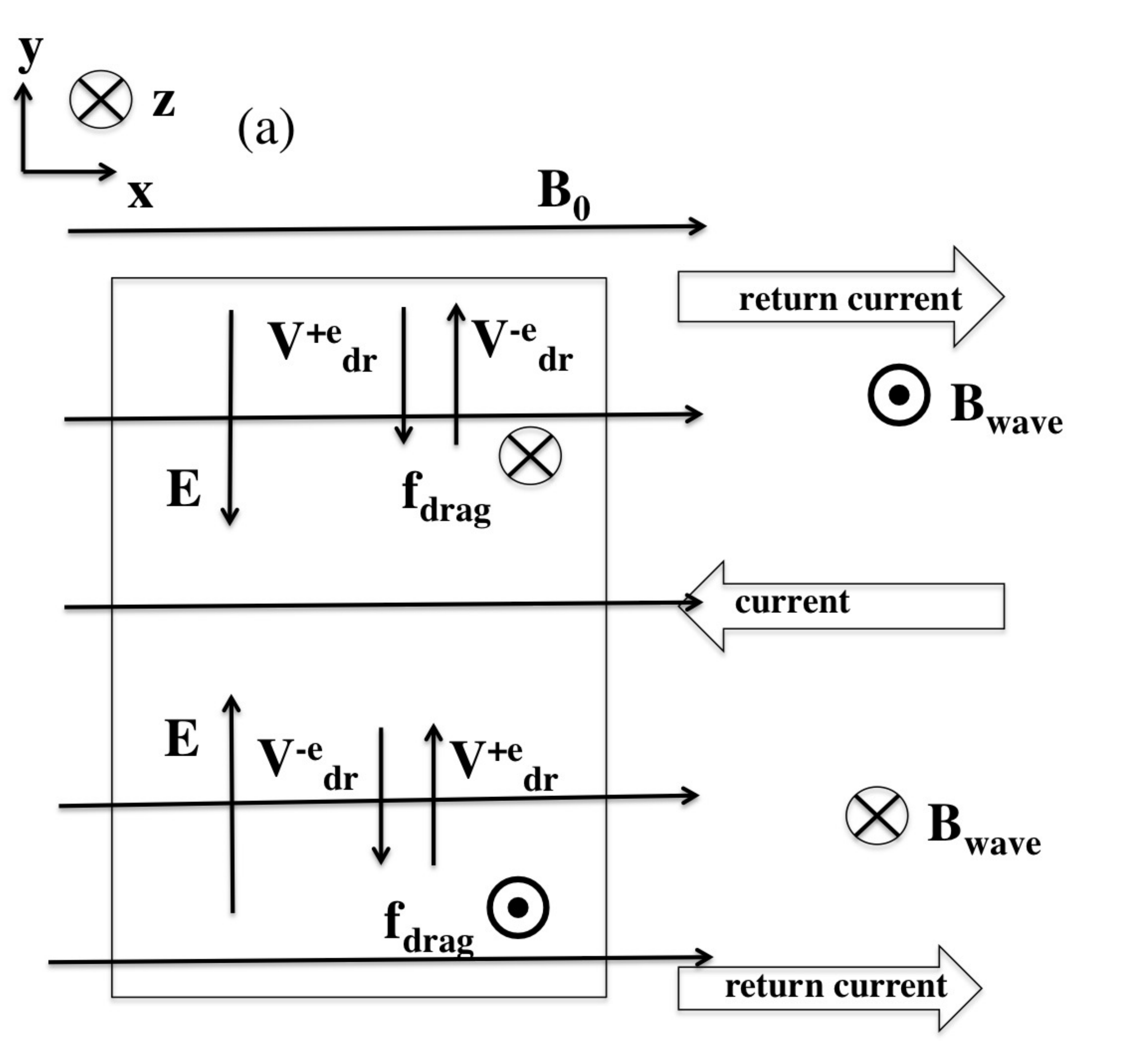}}
{
\includegraphics[width=0.43\textwidth]{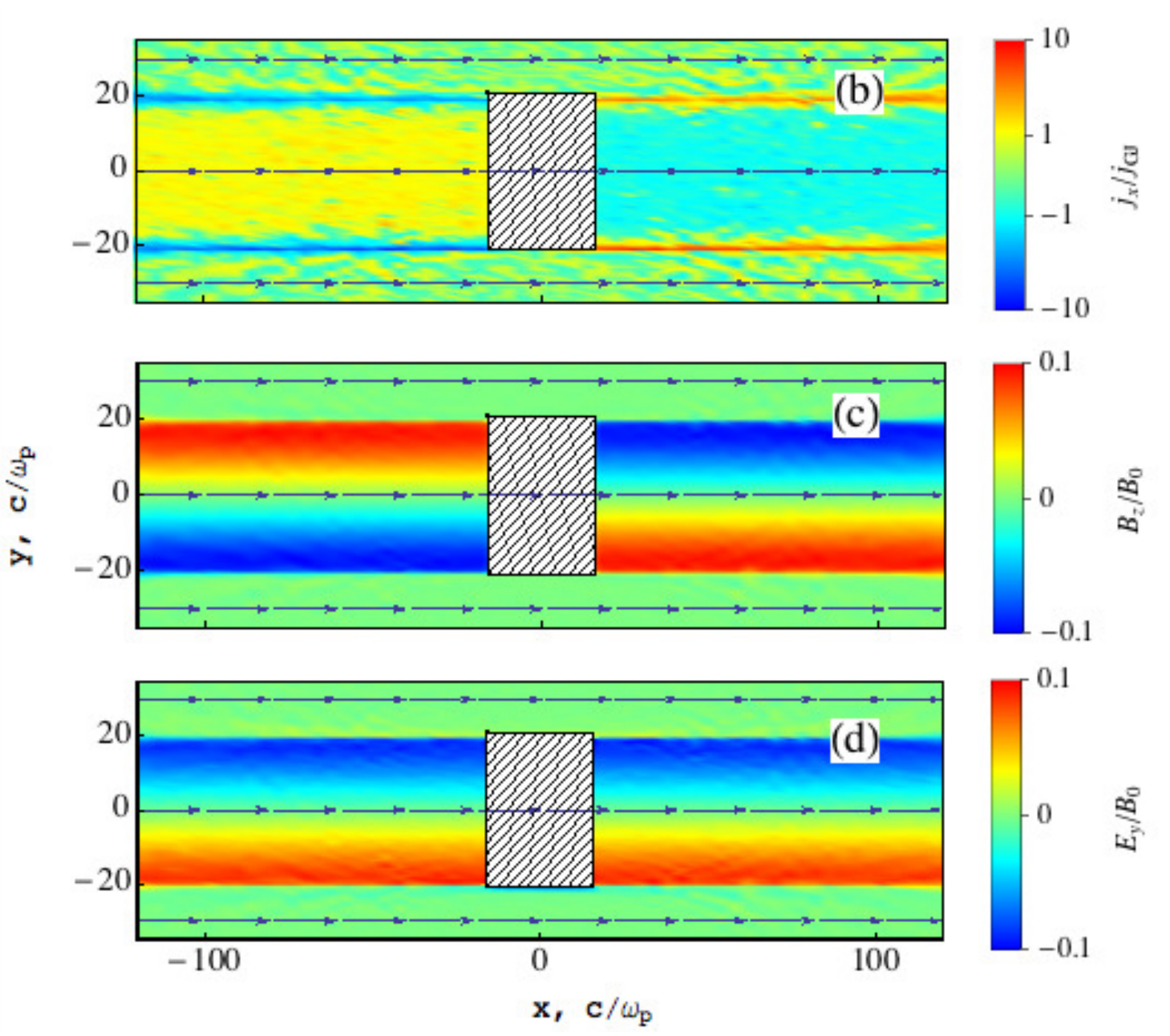}}

\caption{(a): Geometry of the Alfven wave test problem. Applied drag force in
  background magnetic field causes charge separation and gives rise to
the electric field inside the conductor.  (b) The current structure of numerical solution. Alfven wave is launched from the conductor which establishes Goldreich-Julian
current in the bulk
outflow. Thin layers of strong return current at the edges prevent the
conductor from charging. (c) Out of plane component of the magnetic
field $B_z$, normalized by the background field $B_0$. (d) Component
of the electric field $E_y$ tangential to the conductor surface. The
color table is linear in field plots.}
\label{fig1:fluxtube}
\end{figure*}
\label{sec:numerical-models}
To simulate the pulsar magnetosphere we use the 3D electromagnetic PIC code
TRISTAN-MP \citep{anatolycode}. To a good accuracy, the star
can be described as a rotating magnetized conductor. Unipolar induction
generates electric field corresponding to a quadrupolar surface charge. In the limit of zero work function these
charges can be pulled from the surface by the electric field, thus populating the
magnetosphere with plasma. This space-charge limited outflow from a spherical star is difficult to simulate without encountering virtual cathode oscillations and stair-stepping artifacts on a Cartesian grid (\citealt{Anatolychsep}; hereafter, SA02). Instead of prescribing corotating electric fields inside a stair-stepped star and injecting a fraction of the local surface charge at every timestep, we tried to develop a more physical boundary condition for simulating unipolar induction.  
We model the star as a dense magnetized plasma ball with particles that are pushed into rotation by an external drag force. This force emulates the ``lattice" force acting inside a moving solid:  
\begin{equation}
{\bf f_{drag}} = -m ({\bf V} - {\bf V_0})/\tau,
\end{equation}
where $m$ is the particle mass, ${\bf V}$ is the particle velocity,
${\bf V_0}$ is the particle velocity in the steady state and $\tau$ is the equilibration
time. As they are pushed, particles will move across the background magnetic
field ${\bf B}_0$ with the drift velocity
\begin{equation}
{\bf V_{drift}} = c \frac{{\bf f_{drag}}\times {\bf B_0}}{q B_0^2},
\end{equation}
where $q$ is the particle charge. Since the drift velocity depends on particle charge, the drag force will create charge
separation and the corresponding electric field inside the plasma conductor:
\begin{equation}
{\bf E} =  - {\bf V_0} \times {\bf B_0}/c.
\end{equation}
The exterior electric fields will then develop field-aligned components which can extract the surface charges near the edge of the sphere. We find that with this setup the simulation of space-charge limited emission from our plasma conductor does not require special control of the injection rate at the surface. Stair-stepping is also avoided because we do not set boundary fields directly.    
When particles fall back on the conductor, they
are slowed down and forced to move with the star by the
lattice force. 

In the next section, we test our boundary condition by launching an Alfven wave from the surface of a moving conductor.
This process is essential for emitting and returning 
poloidal currents in the rotating magnetosphere that will be described
in section \ref{sec:pulsar}.

\subsection{Test problem: two-dimensional flux tube}
\label{sec:flux tube}

We consider a two-dimensional
magnetized conducting plate of finite thickness that is immersed in background plasma
(``flux tube'', see Fig.~\ref{fig1:fluxtube}a for the description of simulation geometry),
with the following velocity profile inside the conductor: 
\begin{equation}
V_z (y)=  V_0 \frac{y-y_0}{L/2},
\end{equation}
where $y_0$ is the position of conductor's center and $L$ is its
length. The electric field inside the conductor is then
\begin{equation}
E_y (y)=  -V_0 B_0 \frac{y-y_0}{cL/2}.
\end{equation}
Since electric field parallel to the boundary should be continuous,
the same field is established behind the outgoing Alfven wave. When the magnetization is large (Alfven velocity $\approx$ c) toroidal field equals electric field:
\begin{equation}
B_z (y) = E_y(y) =  -V_0 B_0 \frac{y-y_0}{cL/2}.
\label{eq1}
\end{equation}
The charge and current necessary to support this outflow are
\begin{eqnarray}
\rho = \frac{\nabla \cdot {\bf E}}{4\pi} = -\frac{V_0 B_0} {2\pi cL}, \\
{\bf j} = \frac{c \nabla \times {\bf B}}{4\pi} = -\frac{V_0 B_0} {2\pi L} = \rho c, 
\end{eqnarray}
which are just the Goldreich-Julian (GJ) charge and current densities \citep{GJ69}. 

Having these simple analytical results we can check the results of numerical
simulations. The current flow structure is shown in Fig.~\ref{fig1:fluxtube}b. In the bulk outflow the Alfven wave
establishes the GJ current density. We find that the necessary current is launched if plasma
density inside the conductor exceeds the background density $n$ by a factor of 40 for
external magnetization $\sigma\equiv B_0^2/(4\pi n m_e c^2) =  20$ and
particle velocity $V_0 = 0.1 c$ at the edges of plasma plate. The current closure happens in thin return current layers within several
plasma skin depths at the edges of the outflow.
The out-of-plane component of the magnetic field, $B_z$, and the electric field component
 parallel to the conductor surface, $E_y$, are
shown on Fig.~\ref{fig1:fluxtube}(c) and (d). Their profiles agree
with analytical expectation (\ref{eq1}). The ``plasma conductor" boundary condition provides good current closure, and supports the formation of thin current sheets. Despite the current being equal to $\rho c$, the particles in the bulk flow move much slower than $c$ (at most $0.2 c$), because the quasineutral external plasma supports counter-streaming. 

\section{Pulsar magnetosphere}
\label{sec:pulsar}

We now apply the new boundary condition to the
case of 3D magnetized rotating sphere.
We start from a dense pair plasma sphere with radius $R_* = 50$ cells,
located in the center of uniform Cartesian grid with $800^3$ cells. Inside the sphere there are 100
particles per cell, and the local skin depth is only 0.5 cells.  At $t=0$ we
start spinning up the star particles so that after $50$ timest.png particle velocity at the surface is $\Omega_* R_* =
0.375c$ (here $\Omega_*$ is the stellar angular frequency). This defines the size of the
light cylinder $R_{*}/R_{LC} = \Omega_* R_*/c = 0.375$. Such a large ratio of 
$R_{*}/R_{LC}$ is used to minimize the computational expense. We run our
simulations for 7 turns of the star.  The magnetic field is
represented as ${{\bf {B}}_{vacuum}}+{{\bf {B}}_{plasma}}$, where
${{\bf {B}}_{vacuum}}$ is the nonevolving field of the vacuum rotator (we chose it as the sum of uniform magnetic field inside the star and dipole field  
outside with magnetic moment $\mu$), and ${\bf {B}}_{plasma}$ is the
field computed from the currents deposited on the grid (see SA02); ${\bf {B}}_{plasma}$ and the electric field are
set to zero in the beginning. The outer walls of the domain have radiation boundary conditions for both fields
and particles. 

\begin{figure}
\begin{center}
    \includegraphics[width=1\columnwidth]{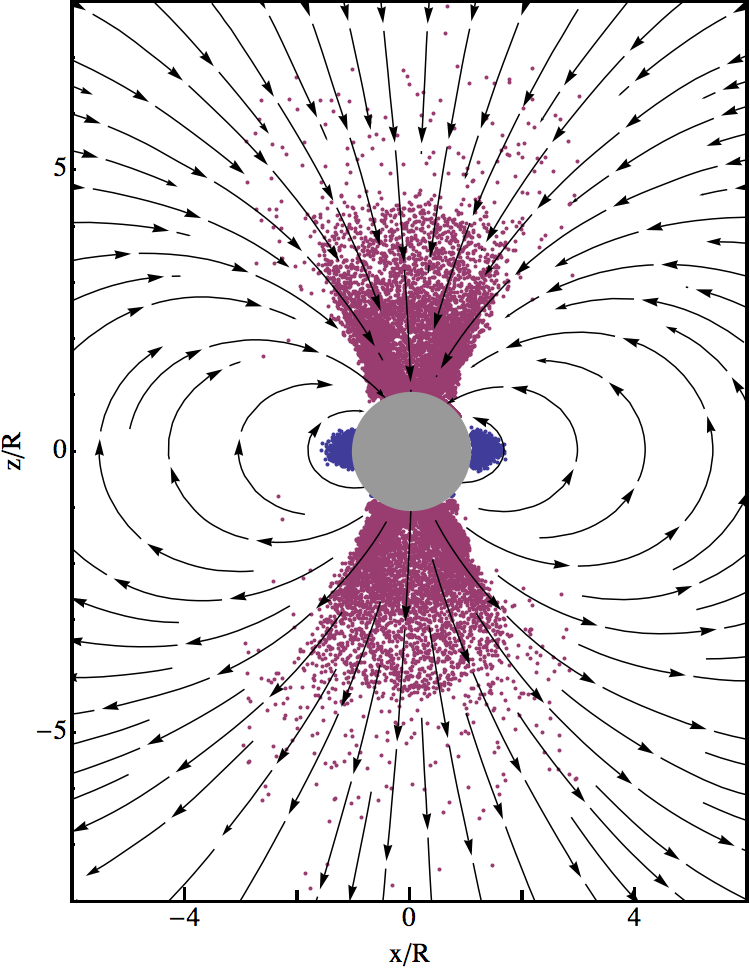}
  \end{center}
  \caption{Slice through 3D charge separated magnetospheric solution
    showing disk-dome configuration. The magnetic field (arrows) does not
    change from the initial dipole outside of the star. Electrons are blue, positrons are red.}
\label{fig:dome}
\end{figure}

\begin{figure}
\begin{center}
    \includegraphics[width=1\columnwidth]{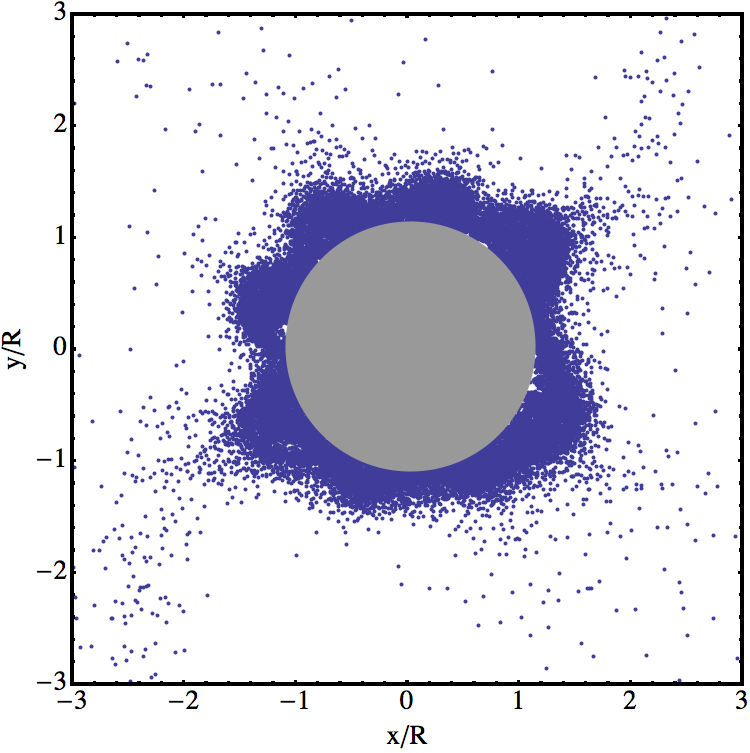}
  \end{center}
  \caption{Diocotron instability in the equatorial plane of
    charge-separated aligned rotator. The
    instability leads to nonaxisymmetric charge modulations and radial expansion of the disk.}
\label{fig:diocotron}
\end{figure}

We discuss two different limits
of plasma supply in the magnetosphere. In the first one, the only source
of plasma particles is the extraction of charges from the central
star; in the second, in addition to the plasma from the star, a
neutral pair plasma is injected throughout the magnetosphere. These limits should capture the transition from ``dead pulsars" with no pair formation to ``active pulsars" with abundant pair formation. The initial evolution is similar for the two cases.  
The unipolar induction generates quadrupolar
electric field, that pulls positive charges near the poles (in our
case rotational and magnetic axes are antiparallel)
and negative charges in the equatorial region, creating a 
charge-separated magnetosphere\footnote{We assume that positive charges are positrons and ignore potential complications of extracting positive charges from the surface of a star.}. If there is no pair formation, the
solution reaches a disk-dome configuration shown in Fig.~\ref{fig:dome} (\citealt{Michel85, Smith}; SA02). 
The dome particles are electrostatically trapped in the vicinity of the star and are rotating with it. 
The equatorial disk is not in corotation with the star and is unstable 
to the diocotron instability (SA02; \citealt{Petri02}; see Fig.~\ref{fig:diocotron} for an equatorial slice of 3D simulation). The instability leads to nonaxisymmetric charge modulations
and radial charge transport in the disk. At the end of our simulation, while some equatorial charges reach the light cylinder (note the extended spiral structures in Fig.~\ref{fig:diocotron}), the current flow is insufficient to appreciably modify the dipole magnetic field structure or cause pulsar spin-down. While $E\cdot B=0$ on the star in this solution, large accelerating gaps exist in the vacuum regions between the domes and the torus. However, $E<B$ everywhere. 

\begin{figure*}
\begin{center}
    \includegraphics[width=2\columnwidth]{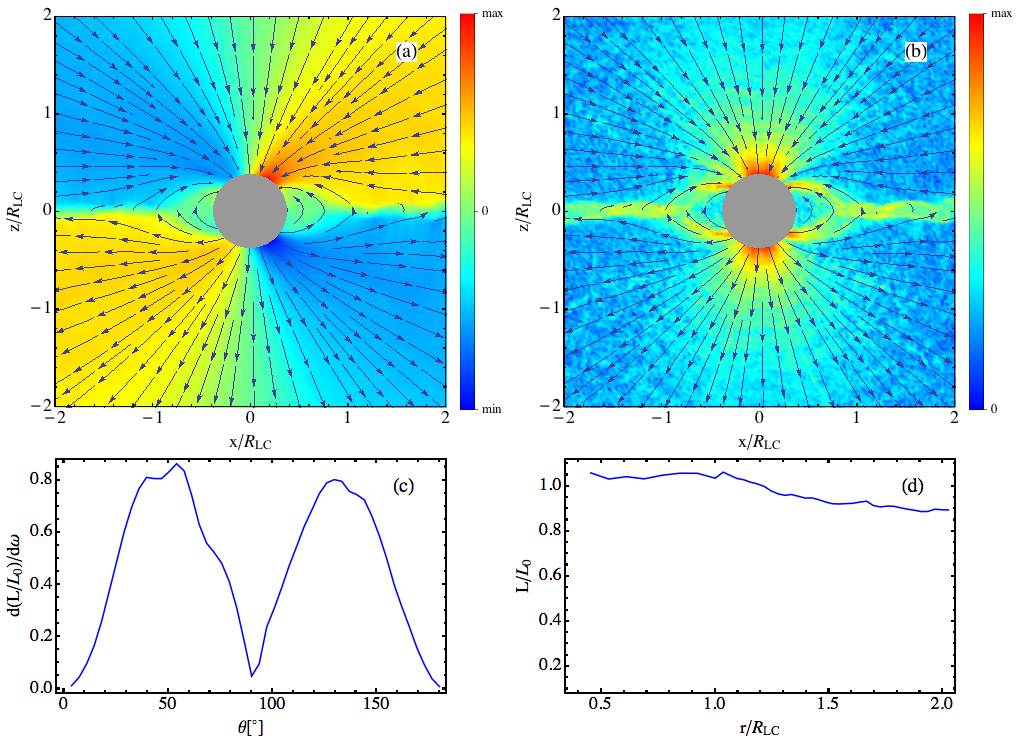}
  \end{center}
  \caption{Slice through 3D aligned pulsar magnetosphere. {(a)}: Poloidal field lines of the
    solution. Color represents the out of plane magnetic field component $B_y$.  {(b)}: Poloidal current flow. Color gives
  $|\sqrt{J^2_x+J^2_z}|$. {(c):} angular distribution of Poynting
    flux per unit solid angle, ${\rm d}L/{\rm d}\omega$, as measured
    at $2R_{LC}$. Normalization constant
    $L_0 = \mu^2 \Omega^4/c^3$ is the energy loss of the
    ideal force-free aligned rotator. The drop at $\theta = 90^{\circ}$ corresponds to the current
    sheet. {(d)}: Poynting
    flux integrated over the sphere as a function of radius from the star. The dissipation is
    less than 15 \% within $2R_{LC}$.}
\label{fig:Magnetosphere}
\end{figure*}

In the second plasma supply scheme, 
initially there is no plasma
outside the star, but at each timestep we inject
neutral pair plasma at a fixed rate in every cell where the magnetization exceeds a certain limit, $\sigma > 400 (R_*/r)^3$.
This limiter helps to prevent overloading of the magnetosphere with plasma,
particularly in the closed zone, while keeping the flow
well-magnetized. The rate of injection over the whole space is 5 GJ
charges per rotation, where we define GJ charge as the integral of GJ
density over a dipolar field inside the light cylinder, $0.8 \Omega
B_0 R^3_* \ln({c/\Omega R_*)}/ ce $. Before the limiter is engaged,
the injected plasma is distributed uniformly over the box, but in
steady state the effective injection falls off as $1/r^3$, and is
mostly important within the light cylinder. The rate of injection in
steady state is $4\, n_{GJ}$ per rotation near the pole, and
$2\, n_{GJ}$  per rotation at the light cylinder, where $n_{GJ}$ is the local GJ density.
In the steady state the number of particles outside the star is constant, and the final magnetization of our solution is $\sigma\approx 400$ near the pole (around 15 particles per cell, local skin
depth is 1.3 cells) and $\sigma\approx 20$ at the light cylinder (around 1 particle per cell, local skin
depth is 5 cells). In steady state the number of simulation
particles is of order $10^8$.
 Magnetization remains nearly constant 
in the wind zone, so the solution is close to force-free conditions everywhere. 
When the magnetosphere is being filled with 
neutral plasma, the charge-separated dome-torus stage is only a 
transient. Two torsional Alfven waves are emitted from
the surface of the star and modify the magnetospheric structure. Near the star the
waves launched from different hemispheres cancel, creating a
region of zero poloidal current. At larger distances along the equator the
current sheet starts to form. While in the beginning the sheet is
charge-separated, the additional neutral plasma inflowing into the
sheet quickly makes it quasineutral, though with a net negative charge (consistent with ${\bf \Omega_*} \cdot
  {\bf B} <0$). The physical thickness of the sheet decreases with additional plasma density, remaining at several local skin depths. 
 
\begin{figure}
\begin{center}
    \includegraphics[width=\columnwidth]{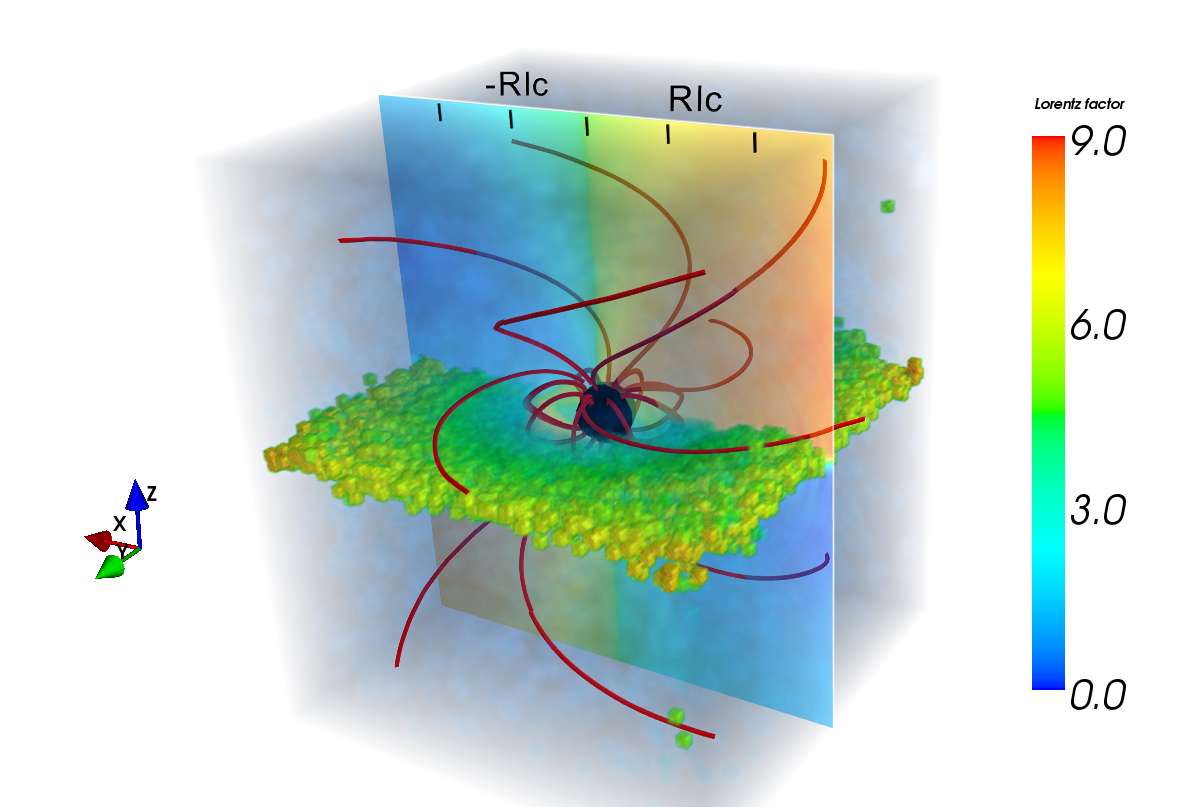}
  \end{center}
  \caption{3D structure of aligned pulsar magnetosphere. Thick lines
    represent the magnetic field lines. Mean Lorentz factor of particles averaged on a coarse grid is shown as 3D volume rendering. The solution does
    not have any strong accelerating regions except for the current sheet where the most
  energetic particle have energies up to $\gamma=15$. Color in the
  plane shows $B_y$ component of the magnetic field. The colorbar for $B_y$ is
  not shown and is similar to Fig. 4a.}
\label{fig:lorentz}
\end{figure}

Magnetic field in a slice through the 3D magnetosphere is shown on
Fig.~\ref{fig:Magnetosphere}a, where color represents the out of plane magnetic
field component $B_y$. The solution is remarkably similar
to force-free or MHD solutions. It consists of the region of the
closed field lines that carries no poloidal current and the region of
open field lines with asymptotically radial poloidal field
lines. The Y-point is located approximately at the light cylinder. A small fraction of field lines close through
the current sheet, suggesting that reconnection occurs there (3D
structure of the field lines is shown in Fig.~\ref{fig:lorentz}). The
poloidal current flow structure is shown with color
in Fig.~\ref{fig:Magnetosphere}b. The current
launched from the polar cap region returns to the star through the current sheet and the
separatrix current layer, which are resolved in our simulation.

The spin-down energy loss of the PIC solution, measured as
Poynting flux integrated over a sphere with $r=R_{LC}$, is $L= (1 \pm 0.1) \mu^2 \Omega^4_{*}/c^3$,
consistent with previous ideal force-free and MHD
work. Fig.~\ref{fig:Magnetosphere}c shows the
angular distribution of the Poynting flux, consistent with the two-peak
structure first found by \citet{tim06}. The radial dependence of the
Poynting flux is shown in Fig.~\ref{fig:Magnetosphere}d. We find that less than 15\% 
of the Poynting flux is dissipated within $2R_{LC}$, in
agreement with \cite{SashaMHD}. However, this differs from the result
obtained by  
\citet{Contopouloslast}, who suggest that
a significant fraction of the Poynting flux is dissipated near the light
cylinder. 
The discrepancy is likely due to the assumption
of null current flowing in the current sheet that was made in
\citet{Contopouloslast}. Plasma in the current sheet beyond the light
cylinder supports counter-streaming and produces space-like current, 
in agreement with the ideal force-free solution \citep{Anatolygamma}.

Our kinetic solution has multiplicity $n/n_{GJ} \approx 10$ near the
pole, so we do not observe significant gap regions where parallel electric field can accelerate particles.
The bulk of the polar outflow 
is thus subrelativistic inside the light cylinder. Nevertheless, particle acceleration is
possible in the equatorial current sheet. 
The spatial distribution of mean particle Lorentz factors is
shown in Fig.~\ref{fig:lorentz}. Energetic particles exist only in
the sheet, reaching maximum energy of $\gamma \approx 15$ for the
solution with magnetization about 20 at the light cylinder. We note that
solutions with smaller magnetization have lower energy of accelerated
particles, $\gamma_{\rm max} \propto \sigma$. The average energy
increases with distance along the sheet, and the spectrum is
consistent with a drifting Maxwellian. Particles are confined in the
sheet and have similar
drift velocities in azimuthal and radial direction. The energy gain is likely due
to heating from reconnection. 
Further acceleration may be possible form the growth of the tearing
mode in the sheet and energization at X-points. Our present runtime
and resolution (2 cells per skin depth in the sheet) are not
sufficient to see this effect yet. The skin depth becomes poorly resolved with time because
of the density increase from plasma inflow into the sheet.
In future work we will study the acceleration mechanism and particle
spectrum in more detail. At five stellar rotational periods we observe the growth of drift-kink instability in the current sheet. 
It causes oscillations of the sheet in the direction
perpendicular to the current, with an amplitude of about several local
plasma skin depths (see Fig.~\ref{fig:Magnetosphere}b). This oscillation widens the current sheet but does not seem to significantly affect the global
magnetospheric structure by the end of our run.

\section{Discussion}

We performed first-principles relativistic PIC simulations of aligned pulsar
magnetosphere by allowing free escape of particles from the stellar
surface and feeding the magnetosphere with neutral plasma.
We confirm that given sufficient plasma supply the magnetosphere
reaches a solution close to the ideal force-free state. The particle
energization in our solution happens only inside the equatorial
current sheet, that is obtained
self-consistently as part of the global magnetosphere structure. The
availability of self-consistent three-dimensional kinetic simulations of
magnetospheres of magnetized conductors will help to develop 
quantitative models of observed radiation from pulsars and other magnetized objects.

Admittedly, our kinetic solution is highly idealized. 
For simplicity we assumed a volume production of plasma, so in the future we will need to relax this assumption and study how realistic pair creation
prescriptions may affect the magnetospheric structure. The reemergence of accelerating regions may be particularly interesting for high energy emission modeling. 
In addition, we will need to extend this study to the
case of oblique rotators. 
Runs with higher resolution will allow us to study the current sheet physics and counterstreaming instabilities more reliably. 
However, even at this stage our kinetic solution is likely a good approximation to the magnetospheric structure of young pulsars with vigorous pair formation.  
More work can also be done on the numerical side. Our boundary condition currently cannot handle 
non-uniform fields inside the star, necessitating a switch to a constant interior field. Although we do not see any adverse effects of this, it will be improved on in
future work. 

Previously, several groups have attempted particle simulations of pulsar magnetospheres with pair production. In SA02, PIC simulations with volume injection of plasma lead to effective reduction of accelerating fields and formation of a wind-like outflow for oblique rotators. However, this failed to work for aligned rotators, due to inconsistent boundary conditions on the star. In the present work, we have rectified this deficiency with our new plasma conductor boundary condition. Aligned rotators were also simulated by \cite{Wada}, using an electrostatic PIC scheme, and by \cite{Gruzinov13_parts}, using PIC with modified particle equations of motion. These authors find solutions with large vacuum gaps and regions with $E>B$, which  accelerate charges to a large fraction of the available potential. The accelerated particles then effectively decouple from the field beyond the light cylinder. We believe such solutions are an artifact of low pair multiplicity in the magnetosphere. Realistic pulsars are expected to effectively produce quasineutral plasma with densities well in excess of local GJ density. Our kinetic solution shows that for pulsars that have large plasma multiplicity there are no pathologies at the light cylinder and any accelerating fields are likely confined to small volumes in the magnetosphere. Also, we find little evidence for dramatic dissipation of Poynting flux beyond the light cylinder (unlike \citealt{Gruzinov12} and \citealt{Contopouloslast}), a conclusion that is likely to only strengthen with improved current sheet resolution. It remains possible that pulsars with weak pair formation form a different class of magnetospheres \citep{Gruzinov13}; however, in the limit of vanishing pair formation these solutions must be bounded by the completely charge-separated solutions that we find here. We conjecture that solutions with large vacuum gaps and pair formation confined to near the star eventually evolve to quasi-neutral configurations, although the time it takes to approach this state may scale inversely with the pair production rate. We will check this conjecture with future simulations. 

We thank Jonathan Arons, Vasily Beskin, Andrei Gruzinov, and Alexander Tchekhovskoy for fruitful discussions. This research was supported by NASA grants NNX12AD01G and NNX13AO80G, Simons Foundation grant 267233, and was facilitated by Max Planck/Princeton Center for Plasma Physics. 
Simulations presented in this article used computational resources supported by PICSciE-OIT TIGRESS High Performance Computing Center.

\end{document}